\documentclass{aa}
\usepackage{natbib}
\bibpunct{(}{)}{;}{a}{}{,} 
\usepackage{graphicx}
\usepackage{txfonts}
\usepackage[bookmarks]{hyperref}

\begin{document}

\title{The role of metallicity in high mass X-ray binaries in galaxy formation models}

\author{M. C. Artale\inst{1,2}, L. J. Pellizza\inst{1,2,3} \& P. B. Tissera\inst{1,2,4}}

\institute{Instituto de Astronom\'{\i}a y F\'{\i}sica del Espacio, C.C. 67,
Suc. 28, (1428) Buenos Aires, Argentina. \and Consejo Nacional de
Investigaciones Cient\'{\i}ficas y Te\'cnicas (CONICET), Argentina 
\and Instituto Argentino de Radioastronom\'{\i}a, Camino Gral. Belgrano km 40, Berazategui, Prov. de Buenos Aires, Argentina
\and Departamento de Ciencias Fisicas, Universidad Andres Bello, Av. Republica 220, Santiago, Chile. }

\offprints{M. C. Artale, \email{mcartale@iafe.uba.ar}}

\date{Received / Accepted}

\abstract{
Recent theoretical works claim that high-mass X-ray binaries could have been important sources of 
energy feedback into the interstellar and intergalactic media, playing a major role in both
the early stages of galaxy formation and the physical state of the intergalactic medium during
the reionization epoch. A metallicity dependence of the production rate or
luminosity of the sources is a key ingredient generally assumed but
not yet probed.}
{Our goal is to  explore the relation between the X-ray luminosity and star formation rate of galaxies as a 
possible tracer of  a metallicity dependence of the production rates
and/or X-ray  luminosities   of high-mass X-ray binaries, using
hydrodynamical cosmological simulations.
}{We developed a model  to estimate  the X-ray luminosities of star
  forming galaxies based on stellar evolution models which include
  metallicity dependences. We applied our
 X-ray binary  models to galaxies selected from 
hydrodynamical cosmological simulations which include chemical
evolution of the stellar populations in 
a self-consistent way. Hence for each simulated galaxies we have
  a distribution of stellar populations with different ages and
  chemical abundances, determined by its formation  history.
This allows us to robustely  predict the X-ray luminosity -- star formation rate
relation under different hypotheses for the effects of metallicity.}
{ Our models successfully reproduce the dispersion in the observed relations as an
outcome of the combined effects of the mixture of stellar populations with heterogeneous
chemical abundances and the metallicity dependence of the X-ray sources.  
We find that the evolution of the X-ray luminosity as a
  function of the star formation rate of galaxies  could store
  information on possible  metallicity dependences of the  high-mass
  X-ray sources. A non-metallicity dependent model predicts a non-evolving
  relation while  any metallicity dependence should affect the
  slope and  the dispersion as a function of redshift. Our results
  suggest the characteristics of the X-ray luminosity evolution
  can be linked to the nature of the metallicity
  dependence of the production rate or the X-ray
  luminosity of the stellar sources. By confronting our models with current available
  observations of strong star-forming galaxies, we find that only
  chemistry-dependent models reproduce the observed trend  for $z <
  4$. However,  it is is not possible to prove
  the nature of this dependence yet.
} 
{}

\keywords{X-ray: binaries -- galaxies: abundances, evolution}
\titlerunning{High mass X-ray binaries in galaxy formation models}
\authorrunning{Artale et al.}
\maketitle

\section{Introduction}
\label{intro}

High-mass X-ray binaries (HMXBs) are systems composed by a compact object, which can be a neutron star (NS) 
or a black hole (BH), and an early-type star. The compact object accretes mass from its companion star,
converting gravitational into thermal energy, part of which is radiated away in the X-ray band ($\sim 0.1-10\ {\rm keV}$). 
Since the first high-energy observatories ({\em Einstein}, {\em ROSAT}, {\em ASCA}), these sources have been observed 
in the Milky Way as well as in nearby galaxies. In the last decade, the higher angular resolution and sensitivity of
{\em Chandra} and {\em XMM-Newton} allowed these observatories to detect thousands of HMXBs in the local Universe 
\citep[][and references therein]{Grimm2003,Fabbiano2006,Mineo2012}. The relation between HMXBs and massive stars makes
these sources dominate the X-ray luminosity of star-forming galaxies with high specific star formation rate (sSFR).
\citet{Mineo2012} have compiled a large sample of HMXBs in nearby late-type galaxies, for which they 
claim that the contamination by other types of sources (i.e. low-mass X-ray binaries ---LMXBs---, background 
active galactic nuclei) is negligible. Recently, the X-ray emission of a sample of metal-poor blue compact dwarf
galaxies was investigated by \citet{Kaaret2011}, and different authors have studied the properties of X-ray emitting 
star-forming galaxies at high redshift \citep{Cowie2012,Basu2012}. These observations have provided a large amount of 
data on the properties of HMXB populations, and on the relation of these properties to those of the host galaxies.

HMXBs are an important tool to investigate stellar (particularly binary) evolution, and the nature of compact objects. 
They are also potential star-formation tracers due to their relation to massive stars, and they have been proposed as 
important sources of stellar energy feedback into the interstellar and intergalactic media 
\citep{Power2009,Mirabel2011a,Dijkstra2011,Justham2012,Power2013}. One of the key problems to understand these systems,
their evolution, and their influence on the environment, is the dependence of the HMXB production and properties on the 
metallicity of the stellar populations from which they form. Recent stellar evolution models suggest that the number of
BHs and NSs produced by a stellar population depends on its metallicity \citep{Georgy2009}. Binary population synthesis 
models show that also the fraction of these compact objects that end up in binary systems with massive companions should 
depend on metallicity \citep{Belczynski2004a,Dray2006,Belczynski2008,Belczynski2010a,
Linden2010}, because at lower metallicities more systems can survive disruption when the
 primary BH forms, and also avoid
merging in the common-envelope phase. Finally, both models and observations suggest that low-metallicity stars form more 
massive BHs, which could produce potentially higher-luminosity HMXBs \citep[][and references therein]{Belczynski2010b,Linden2010,Feng2011}.

However, the observational evidence for the metallicity dependence of the number and luminosity function of HMXBs is still poor. 
It is clear that this dependence must be searched for in the properties of the populations of HMXBs in star-forming galaxies. 
A key observable is the X-ray luminosity $L_{\rm X}$ of such galaxies, which in the local Universe scales with the star 
formation rate \citep[SFR;][]{Grimm2003,Mineo2012}. This correlation is usually parameterized as
$L_{\rm X} = 3.5 \times 10^{40} {\rm erg}\ {\rm s}^{-1}\ f_{\rm X}\ {\rm SFR}/(M_\odot\ {\rm yr}^{-1})$, 
where the factor $f_{\rm X}$ accounts for possible variations due to the dependence of HMXB properties on metallicity
or other physical parameters. \citet{Mineo2012} found that observations of nearby galaxies are consistent with a constant
$f_{\rm X} \sim 0.2$, but the correlation shows a large dispersion, which might be due to metallicity effects. 
\citet{Kaaret2011} measured unusually larger $f_{\rm X}$ values for a sample of nearby blue 
compact dwarf galaxies with low metallicities. Unfortunately, their small sample did not allow them to reach statistically
meaningful conclusions about a departure of these galaxies from the standard $L_{\rm X}$--SFR relation of \citet{Grimm2003}.
\citet{Cowie2012} investigated this issue using a sample of galaxies at high redshift, for which metallicity effects should be 
important due to the chemical evolution of the Universe. They found that $f_{\rm X}$ is at most marginally dependent on redshift, 
however the observational uncertainties and the complex dependence of galaxy metallicity on redshift still leave the question open.
Using the same X-ray survey \citet{Basu2012} have studied Lyman-Break galaxies in the range $z=1.5-8$, finding instead that $f_{\rm X}$
evolves with redshift. They also argue that \citet{Cowie2012} did not correct the galaxy luminosities for dust attenuation, which could 
prevent them to observe the evolution. A key issue to resolve the problem is to understand how $f_{\rm X}$ is 
affected by the metallicity dispersion within a galaxy, the correlation of the galaxy mean metallicity and SFR, and the chemical 
evolution of the Universe, in order to make a proper interpretation of the observational results.

An interesting way to explore the metallicity dependence of HMXB populations, and the key issue of the possible evolution 
of $f_{\rm X}$, is through the combination of binary population synthesis models with a description of the stellar populations 
in a galaxy. \citet{Belczynski2004b} used this method to develop models that reproduce the emission of specific galaxies, 
while \citet{Zuo2011} explored the X-ray emission of galaxies at different redshifts using prescriptions for their star formation histories.
These authors found a good agreement between the predicted and observed X-ray luminosity to stellar mass ratio in the range $z = 0-4$, 
but they were not able to reproduce the corresponding X-ray to optical luminosity ratio. A step forward in this 
approach is to couple binary population synthesis models to scenarios for the formation and evolution of galaxies 
in a cosmological context. Previous works used population
  synthesis models which provide a description of the HMXB properties expected from 
a single, homogeneous parent stellar 
population or semi-analytical models where there is a unique mean
metallicity for stellar populations born at a certain time in a given
galaxy. As we mentioned before, we intent to improve the modeling of
HMXBs by providing a  more realistic description of the complexity of
stellar populations within a galaxy.  

Here we present a novel scheme to model the HMXB populations of star-forming galaxies, which couples population synthesis
results to galaxy catalogues constructed from a hydrodynamical cosmological simulation of structure formation which is part
of the Fenix project (Tissera et~al., in prep.). This simulation includes star formation, a multiphase treatment of
the interstellar medium, the chemical enrichment of baryons, and the feedback from supernovae in a self-consistent way
\citep{Scannapieco2005,Scannapieco2006}, and reproduces global dynamical and chemical properties of galaxies 
\citep{deRossi2010,deRossi2012,pedrosa2014}. This makes the simulation well suited for the task of 
investigating the evolution of the $L_{\rm X}$--SFR relation of star forming galaxies, expanding and complementing
the results obtained by other methods such as semi-analytical models \citep[e.g.,][]{Fragos2012}. Our scheme is 
similar to those applied by \citet{Nuza2007}, \citet{Chisari2010}, \citet{Artale2011a}, 
and \citet{Pellizza2012} to the study of gamma-ray bursts. It includes both the modelling of the intrinsic HMXB 
populations of galaxies, and the definition of different samples comparable to observations, based on the modelling 
of selection effects. This is an important ability of our scheme, as it allows us to make a fair comparison with 
observations to constrain free parameters and discard incorrect hypotheses. Using our scheme, we develop different
models to explore the effects of the dependence of the HMXB population properties on the metallicity of the parent 
stellar populations, and address the question of the evolution of the $f_{\rm X}$ factor by comparing our predictions 
to observations of galaxies across time.

This paper is organized as follows. In Section~\ref{simu} we briefly present the numerical simulations used 
to describe the formation and evolution of galaxies, and the construction of galaxy catalogues. 
 In Section~\ref{popsyn} we describe our HMXB model and how it is implemented onto the simulated galaxy catalogues
to generate intrinsic population. 
In Section~\ref{redshift_g}, we present our results as a function of cosmic time. 
Finally, in Section~\ref{con}, we discuss our main conclusions.

\section{Simulations}\label{simu}

The simulation used in this work (S230A) is part of the Fenix project \citep{pedrosa2014}, 
designed to study the role of metals in galaxy formation (Tissera et~al., in prep.). 
The run was performed with a version of {\small GADGET-3} \citep{Springel2005} which 
includes star formation, metal-dependent cooling, chemical enrichment, a multiphase model
for the interstellar medium, and supernovae feedback \citep{Scannapieco2005,Scannapieco2006}.

The feedback model includes Type II and Type Ia supernovae (SNII and SNIa, respectively) 
for chemical and energy production, within  a multiphase model for the 
interstellar medium. The SN feedback model considers as progenitors of SNII  stars
with masses larger than $8\,M_{\odot}$ and assumes  lifetimes of $\approx 10^6\,{\rm yr}$. 
The lifetimes of SNIa are selected randomly within the range $0.1-1\,{\rm Gyr}$. 
A constant ratio between the rates of SNII and SNIa is assumed. In the analysed 
simulation the thermal energy released into the interstellar medium per SN event 
is $0.7 \times 10^{51}\,{\rm erg}$. The chemical algorithm from \citet{Mosconi2001} 
was adapted for GADGET-2 by \citet{Scannapieco2005}, with initial primordial abundances 
for gas particles of $X_{\rm H}=0.76$ and $X_{\rm He}=0.24$. The algorithm follows the
enrichment of 12 isotopes: $^1$H, $^2$He, $^{12}$C, $^{16}$O, $^{24}$Mg, $^{28}$Si, $^{56}$Fe,
$^{14}$N, $^{20}$Ne, $^{32}$S, $^{40}$Ca and $^{62}$Zn. The chemical yields for SNII 
are taken from  \citet{Woosley1995} while for SNIa, we adopt the W7 model of \citet{Thielemann1993}.
The SN feedback model used in S230A has proven to be successful at regulating the star formation 
activity, and at driving powerful mass-loaded galactic winds without the need to introduce mass-dependent 
parameters.

The simulation S230A  describes a cubic volume of $10 h^{-1}\,{\rm Mpc}$ per side, consistent with a 
$\Lambda$-CDM universe with cosmological parameters $\Omega_{\Lambda}=0.7$, $\Omega_{\rm m}=0.3$,
$\Omega_{\rm b}=0.04$, $\sigma_{8}=0.9$, and ${\rm H}_{0}=100 h\,{\rm km} \,{\rm s}^{-1}\,{\rm Mpc}^{-1}$,
where $h=0.7$. Initially the simulation has $2 \times 230^3$ particles in total, with masses of 
$8.47 \times 10^{6}\,M_{\odot}$ for dark matter and $1.3\times 10^{6}\,M_{\odot}$ for gas particles.
S230A has been used by \citet{deRossi2010} and \citet{deRossi2012} 
to study the Tully-Fisher relation, obtaining good agreement with observations. Furthermore, 
the evolution of the dark matter halo mass function as a function of
redshift reproduces that  obtained from the Millenium Simulation
\citep{deRossi2013}. Nevertheless we acknowledge the fact that
  this simulation overproduces the stellar mass formed at high
  redshift which is an ubiquitous problem for $\Lambda$-CDM scenarios. In fact, in a
  forthcoming paper we will explore how HMBXs might contribute to solve this
  problem.

The virialized structures were selected by using a friends-of-friends technique and the 
substructures within their virial radii were identified with the {\small SUBFIND} algorithm
\citep{Springel2001}. Galaxy catalogues were constructed from $z \sim 0$ to $z\sim 5$ by selecting 
those substructures with more than 3000 particles, which is equivalent to have galaxies with 
stellar masses above $\sim 10^8\,M_{\odot}$. For each simulated galaxy we have the chemical 
abundances and ages of its stellar populations. The mass-metallicity relation (MZR) of galaxies
in the simulation was studied by \citet{deRossi2010}, finding a lower mean metallicity for a fixed
stellar mass than that observed by \citet{Tremonti2004}. Hence, we renormalized the simulated 
abundances adopting the results from \citet{Maiolino2008} to make them consistent with current observational results.

\begin{figure}
\centering
\includegraphics[width=0.4\textwidth]{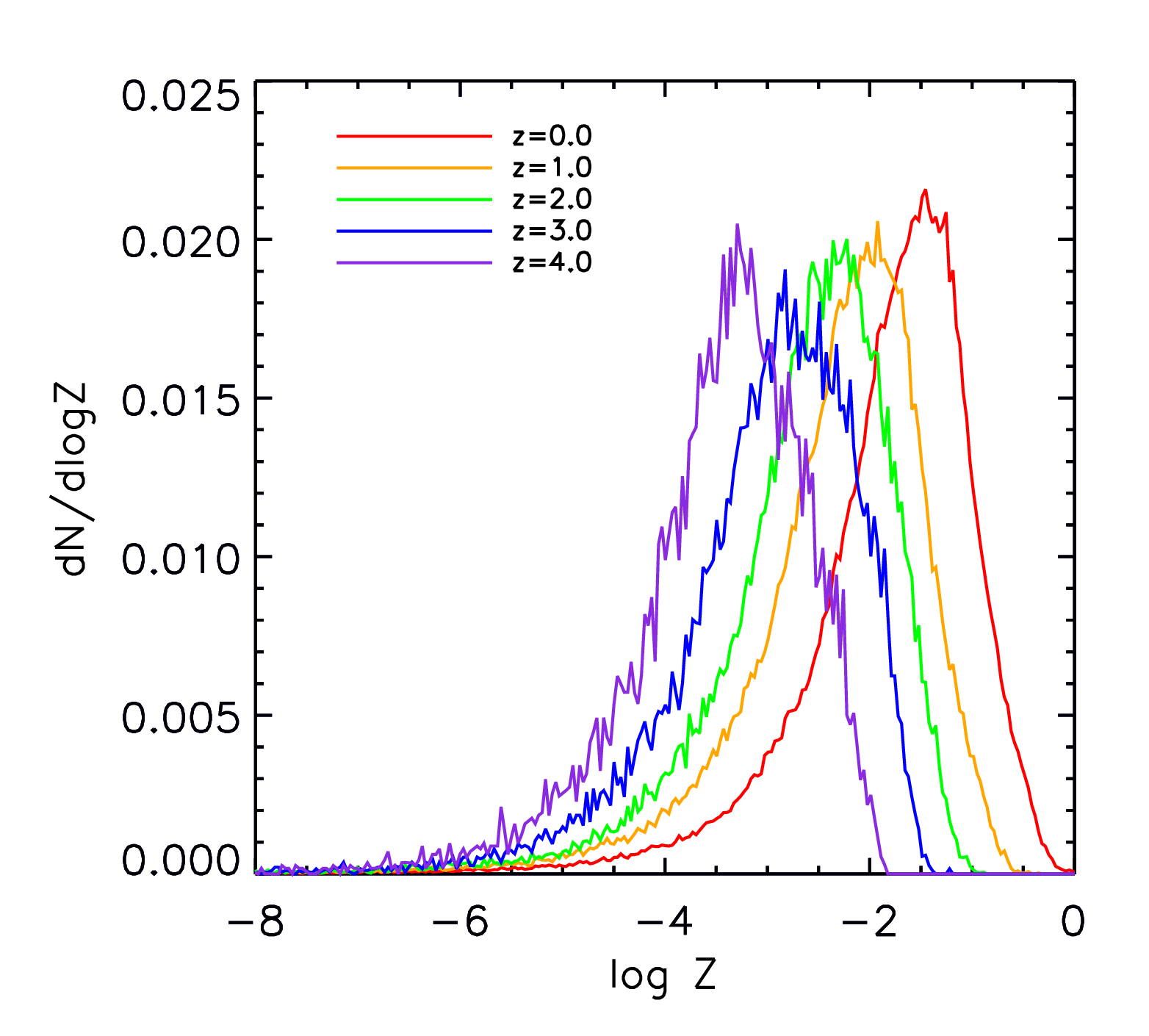}
\includegraphics[width=0.4\textwidth]{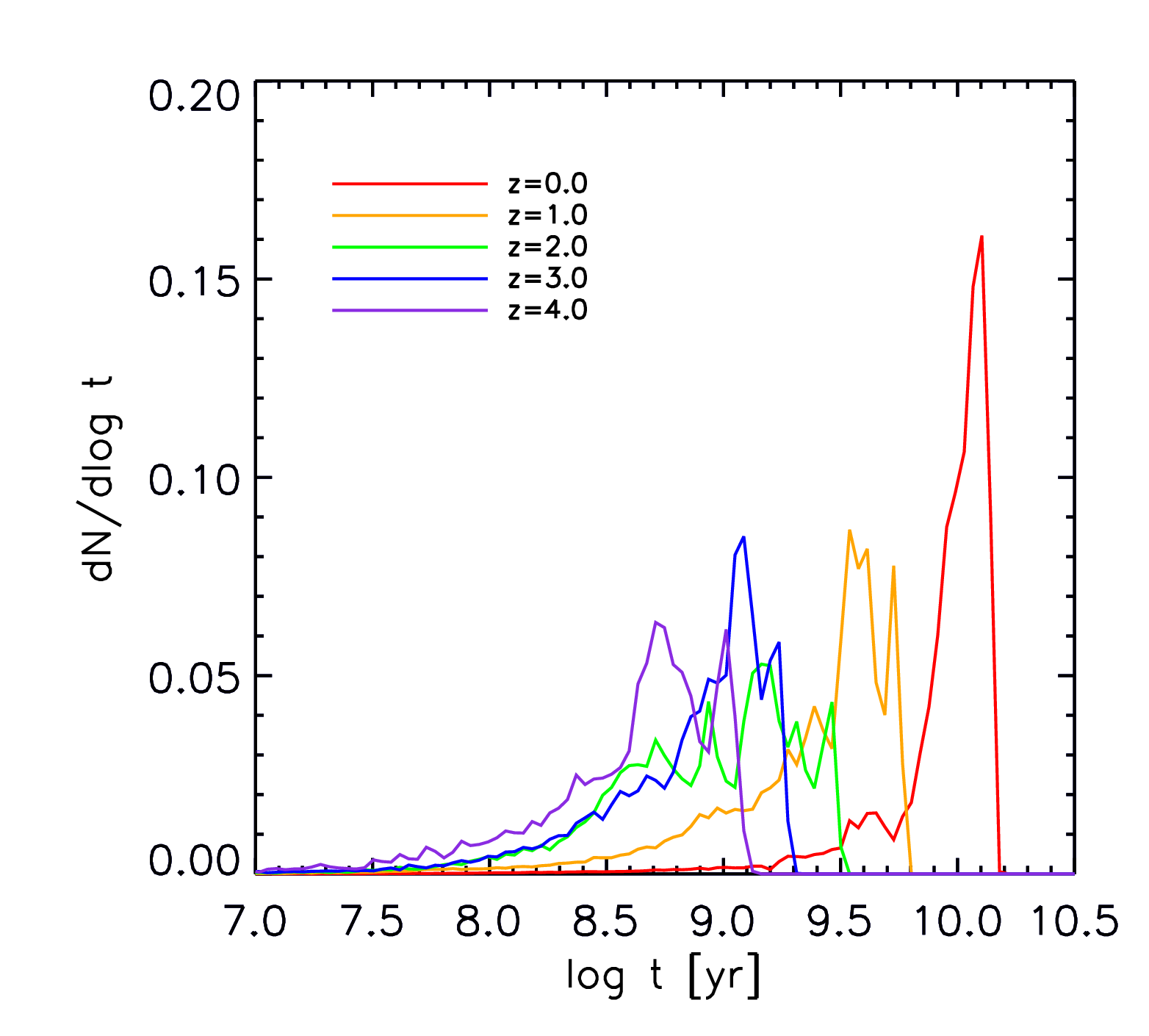}
\caption{
Mass fraction of the stellar populations as a function of  metallicity (Z, upper panel) and age (lower panel) for a
typical galaxy. The distributions are displayed for different redshifts. At each available time, HMXB progenitors are
  selected from these stellar populations by following constrains on
  age and metallicity described in the following sections.}
\label{metages}
\end{figure}

Each of the simulated galaxies is formed by a combination of
  stellar populations spatially distributed according to its formation history.
 The new-born stars have different
  chemical abundances and it is from these mixed stellar populations
  that the progenitors of HMXBs are chosen.
To illustrate this point, in Fig. ~\ref{metages} we show the stellar mass fractions
 as a function of metallicity ($Z$) and of age for
one of the simulated galaxies, for different redshifts.
 Metallicity
is defined as the ratio between the total mass in chemical elements
heavier than He and the stellar mass of a given stellar population. Each
analysed galaxy has more than 1000 star particles which represent
its stellar populations. These distributions vary for galaxies with diferent assembly histories.
This is one of the major advantage of using hydrodynamical cosmological simulations. A drawback of our model it is that we
 have simplified scheme for HMXBs compared to those of
\citet{Fragos2012}, for example. However,  each approach can contribute
with a different insight to a given problem and it is from the
complementary findings that deeper knowledge might be achieved.

\section{Population synthesis} 
\label{popsyn}

\subsection{The HMXB population of a galaxy}
\label{res2}

For each simulated galaxy, we produce a synthetic HMXB population using the 
properties of its stellar populations (i.e., the metallicities and ages of more than a few thousands of particles describing them). 
Given that the progenitors of HMXBs are massive stars, the compact objects
in these systems form shortly after a star formation episode. Both theoretical
models \citep{Belczynski2008} and observations \citep{Shtykovskiy2007} suggest
that the X-ray emission peaks roughly 20--40~Myr after the starburst. 
Hence, we assume that only young stellar populations will contribute
with HMXBs, and select only those stellar populations with ages lower than 100~Myr.
Our results do not depend strongly on the particular choice of this value, as
far as its order of magnitude is preserved.

We compute the number of NS and BHs produced by each stellar population 
selected from the galaxy catalogues, according to the model of \citet{Georgy2009}
for the evolution of massive stars. This model includes stellar rotation and the 
dependence of the type of compact remnant on the metallicity of the progenitor star.
An important feature of this model is the prediction of a higher fraction of BHs as
the metallicity decreases. For a stellar population $p$ with mass $m_{*,p}$ and
metallicity $Z$, formed at redshift $z_p$, we calculate the number of BHs and NS as

\begin{equation}
N^{\rm BH,NS}_p = m_{*,p} \frac{\int_{R^{\rm BH,NS}(Z)} \xi(m) dm}{\int^{100\,M_\odot}_{0.1\,M_{\odot}} m \xi(m) dm},
\end{equation}

\noindent
where $R^{\rm BH,NS}(Z)$ defines the mass range over which each type of 
compact remnant is produced at metallicity $Z$, $\xi(m)$ is the 
\citet{Salpeter1955} initial mass function (IMF) with lower and upper mass 
cut-offs of $0.1\,M_\odot$ and $100\,M_\odot$, respectively. This choice of
IMF is consistent with that implemented in the analysed  numerical cosmological simulation.

A fraction $f_{\rm bin}^{\rm BH,NS}$ of the progenitors of these compact objects belongs primordial binaries, and a fraction
 $f_{\rm rem}^{\rm BH,NS}$ of these binaries remains bound after the supernova event in which the compact objects are produced. 
A fraction $f_{\rm HM}^{\rm BH,NS}$ of these X-ray binaries will have high-mass companions. Finally, only a fraction
 $f_{\rm acc}^{\rm BH,NS}$ of these will be in the X-ray binary phase (i.e., accreting matter from the companion and
 emitting X-rays) at any time. Hence, the ratio of the number of emitting HMXBs to that of the compact objects produced by the stellar population is

\begin{equation}
f_{\rm HMXB}^{\rm BH,NS} = f_{\rm bin}^{\rm BH,NS} f_{\rm rem}^{\rm BH,NS} f_{\rm HM}^{\rm BH,NS} f_{\rm acc}^{\rm BH,NS}.
\label{fracbin}
\end{equation}

\noindent
Then, the number of HMXB systems produced in each stellar population in our galaxy catalogues is

\begin{equation}
N_{{\rm HMXB}, p} = f^{\rm BH}_{\rm HMXB} \, N^{\rm BH}_p + f^{\rm NS}_{\rm HMXB} \, N^{\rm NS}_p,
\label{binnumpart}
\end{equation}

\noindent
and the number of HMXBs in each galaxy $g$ is

\begin{equation}
\label{binnum}
N_{\rm HMXB} = \sum_{p \in g} N_{{\rm HMXB}, p}.
\end{equation}

Theoretical models suggest that the ratio $f^{\rm BH,NS}_{\rm HMXB}$ are time and metallicity dependent. The metallicity dependence arises mainly in 
the corresponding dependence of the mass loss of each star and the mass transfer between them. The time dependence 
is due mainly to two facts: first, many X-ray binaries are transient sources with very short timescales 
(months, years) compared to stellar evolution \citep[e.g.][]{Fabbiano2006}; second, the onset and end of 
the X-ray binary phase depend on the evolutionary paths of both binary components. Eqns.~\ref{binnumpart} 
and \ref{binnum} show that the time and metallicity dependences of the number of HMXBs couple to the age and 
metallicity spreads of the stellar populations within a particular galaxy to produce the final number of HMXBs 
in that galaxy. This is an original feature of our models,  not present in previous works, which used semianalytical 
models for galaxy formation and evolution. The final number of HMXB depends not only on the details of stellar 
evolution, but also on the assembly history of each galaxy. Moreover, the age and metallicity dependences 
produce different effects. While the latter couples to a strongly peaked metallicity distribution, the former 
couples to a slowly varying age distribution (in the age range of interest, 10--100~Myr; Fig.~\ref{metages}).
 This makes the time dependence to be smoothed, as similar numbers of stellar populations of different ages 
contribute to produce the final number of HMXBs. This is not true for the metallicity dependence. Hence, we 
expect metallicity effects to be enhanced with respect to age effects.

Based on the above discussion, we assume as a working hypothesis that age effects can be neglected, replacing 
all time-dependent quantities in Eqn.~\ref{fracbin} by their time averages. Taking advantage of this fact, we 
parameterize the ratio of the number of emitting BH-HMXBs to that of the compact objects produced by the stellar population as

\begin{equation}
f_{\rm HMXB}^{\rm BH} (Z) = \eta f_{\rm B04}^{\rm BH}(Z),
\label{binaryratio}
\end{equation}

\noindent
where $f^{\rm BH}_{\rm B04}$ is the remaining fraction of binary systems with a BH and a non-compact companion at 
any time, given by the population synthesis models of \citet{Belczynski2004b}. We use the values of 
$f^{\rm BH}_{\rm B04}(Z)$ at an age of 11~Myr because only at this age the metallicity dependence is given.
 The free parameter $\eta$ takes into account the correction needed to transform this particular value into
 the time average, the fraction of transient sources, and the fraction of high-mass companions. 
As these three factors are poorly known, we will estimate the value of $\eta$ in Sect.~\ref{free_par} by
 requiring the models to fit the observations of HMXB populations of local galaxies. We will assume then 
that $\eta$ is independent of redshift to predict the properties of HMXB populations at any redshift.


The values of $f^{\rm BH}_{\rm B04}(Z)$ are given for $Z_1 = 0.02$, $Z_2 = 0.001$, and $Z_3 = 0.0001$. 
Hence, we adopt three metallicity bins centred at these values, and assign the corresponding fractions
 $f^{\rm BH}_{\rm HMXB}$ to stars according to their metallicity $Z$ (see Table~\ref{tab_parameters}).
 We adopt a simple model to estimate the  fraction for NS, $f^{\rm NS}_{\rm HMXB}(Z) = f^{\rm BH}_{\rm HMXB}(Z)$.
 Note that the total number of HMXBs in a galaxy depends on the particular distribution of metallicities of 
its stellar populations, and not on its mean metallicity. And since our numerical simulations provide these
 distributions as galaxies evolve, they are expected to describe more realistically the X-ray binary populations.

Given the number of HMXBs in a galaxy, we compute its total X-ray luminosity $L_{\rm X}$ by assuming an X-ray 
luminosity function (XLF) for the HMXBs. For this work, we adopt the XLF estimated from observations of
 HMXBs in the local Universe reported by \citet{Mineo2012}, $\Psi(L_{\rm X}) \propto L_{\rm X}^\alpha$, 
with $\alpha = -1.6$ and $L_{\rm X} \in [L_{\rm min}, L_{\rm max}]$. From the above discussion, it follows 
that the intrinsic X-ray luminosity of a galaxy in our model is given by 

\begin{equation}
L_{\rm X} = \sum_{p \in g} N_{p, {\rm HMXB}}(Z) \langle L_{\rm X}(Z) \rangle.
\label{l_intrinsic}
\end{equation}

To sum up, the free parameters of our HMXB model are $\eta$, $L_{\rm min}$ and $L_{\rm max}$. Only the 
last one is physically interesting, because it is related to the mass distribution of compact objects,
 and to the accretion process. As we want to explore the dependence of $L_{\rm max}$ on metallicity, 
we will create several models by assuming different forms for this dependence. For each model, the 
values of $\eta$ and $L_{\rm min}$ will be determined by requiring the synthetic HMXB populations to 
reproduce available observations of nearby galaxies, as discussed in Section~\ref{free_par}.

\begin{table*}
\centering
\caption{Main parameters of the HMXB models.}
\label{tab_parameters}
\begin{tabular}{cccccccccc}
\hline \hline
Model & $f^{\rm BH}_{\rm b}(Z_1)$\tablefootmark{a} & $f^{\rm BH}_{\rm b}(Z_2)$\tablefootmark{a} & $f^{\rm BH}_{\rm b}(Z_3)$\tablefootmark{a} &
$\alpha$\tablefootmark{b} & $\eta$\tablefootmark{c} & $L_{\rm min}$\tablefootmark{d} & $L_{\rm max} (Z_1)$\tablefootmark{e}
& $L_{\rm max} (Z_2)$ & $L_{\rm max} (Z_3)$  \\
 & & & & & & ($10^{35}{\rm erg\,s}^{-1}$) & (${\rm erg\,s}^{-1}$) & (${\rm erg\,s}^{-1}$) & (${\rm erg\,s}^{-1}$) \\
\hline
M0 & 1.0 & 1.0 & 1.0 & -1.6 & $1.75 \times 10^{-4}$ & $10$  & $10^{40}$ & $10^{40}$ & $10^{40}$  \\ 
M1 & 0.14 & 0.09 & 0.02& -1.6 & $1.75 \times 10^{-2}$ & $1$ & $10^{40}$ & $10^{40}$ & $10^{40}$  \\ 
M2 & 0.14 & 0.09 & 0.02 & -1.6 & $1.75 \times 10^{-2}$ & $1$ & $10^{42}$ & $10^{40}$ & $10^{40}$  \\ 
M3 & 0.14 & 0.09 & 0.02 & -1.6 & $3.00 \times 10^{-3}$ &  $5$ & $10^{42}$ & $10^{40}$ & $10^{40}$ \\ 
\hline
\end{tabular}
\tablefoot{
 The rate of HMXB progenitors and X-ray luminosity do not depend on metallicity in model M0. Model M1 assumes
 that only the rate of binary systems depends on metallicity.
 Models M2 and M3 include both metallicity-dependent rate and X-ray luminosity.\\
\tablefoottext{a}{Fraction of BHs in binary systems. For chemistry-dependent models (M1, M2 and M3) 
we derive the values from \citet{Belczynski2004a}.}
\tablefoottext{b}{Index of the power law XLF from \citet{Mineo2012}.}
\tablefoottext{c}{The normalization factor. This value is a free parameter of the model (see Section \ref{popsyn}) which
is fitted with observations of nearby galaxies.}
\tablefoottext{d}{Minimum X-ray luminosity. This value is a free parameter of the model.}
\tablefoottext{e}{Maximum X-ray luminosity. Models M2 and M3 assume that the maximum X-ray luminosity of HMXBs depend on metallicity.}

}
\end{table*}

Finally, as an {\em a posteriori} check of the consistency of assuming that the time dependence of the
 HMXB number is negligible, we created a new set of models by introducing a time-dependent factor in the 
fraction $f^{\rm BH,NS}_{\rm HMXBs}$. This factor was chosen as the Gaussian function that best fits the 
results of \citep{Shtykovskiy2007} for the relation between the number of sources and their age. 
The new results agree with those of our former models to within a few percent, supporing our assumption.

\subsection{The detectability of HMXB populations}
\label{res_o}

To make a proper comparison with observations, we must model the selection effects
present in the observed samples and  produce an {\em observable sample} of HMXB populations 
from the intrinsic population defined in the previous section.
Observed X-ray luminosities can be the contribution of both  HMXBs and LMXBs. Hence following \citet{Mineo2012}
we only take observed galaxies with ${\rm sSFR} > 10^{-10} {\rm yr}^{-1}$  which are dominated by massive star formation 
and the same  constrain is imposed  on our simulated galaxy sample.

As the selection effects  are different for nearby and high-redshift galaxy samples, we model them separately. 
For nearby galaxies,
satellites can resolve each HMXB as a separate point source if its luminosity is 
above some threshold value $L_{\rm lim}$, which depends on the distance to the source. 
Hence, the observed number of sources and total X-ray luminosity of a galaxy are given by

\begin{equation}
N^{\rm obs}_{\rm XMXB} = \sum_{p \in g} N_{{\rm HMXB}, p}(Z) f_{\rm obs},
\end{equation}

\noindent
and

\begin{equation}
L^{\rm obs}_{\rm X} = \sum_{p \in g} N_{{\rm HMXB,} p}(Z) f_{\rm obs} \langle L_{\rm X} (Z)\rangle_{\rm obs},
\end{equation}

\noindent
where $f_{\rm obs}$ is the fraction of sources with luminosity greater than $L_{\rm lim}$,
and $\langle L_{\rm X}(Z) \rangle_{\rm obs}$ is the mean luminosity of these sources.
The effect of imposing this luminosity threshold  $L_{\rm lim}$ is the decrease of the observed X-ray luminosity
of a given galaxy. 

 At $z = 0$, we confront our models to the observations of \citet{Mineo2012} who
 correct the total X-ray luminosity 
of the population of HMXBs to include unobserved sources down to $L_{\rm lim}=10^{36}\ {\rm erg\, s}^{-1}$.
Hence, we took this value for $L_{\rm lim}$ when computing the X-ray
luminosity of galaxies at $z = 0$ in our model. Note that for consistency, $L_{\rm min}$ should be smaller than  $L_{\rm lim}$. 
As the XLF was derived by these authors using observations 
in the rest-frame $0.5-8\,{\rm keV}$ band, our luminosities refer to
this band. For the number of HMXBs, instead, we use a prescription for $L_{\rm lim}$ obtained from its
correlation with the SFR, observed in the sample of \citet{Mineo2012}.

For high redshift galaxies, satellital instruments can only detect the combined X-ray
luminosity of their HMXB populations. In some cases, individual galaxies are not detected at
all, but stacking their images allows observers to measure the sum of their X-ray luminosities. 
In this way, \citet{Basu2012} obtain an estimate of the mean X-ray luminosity of different samples
of galaxies. In particular, they investigate the luminosity of Lyman-break galaxies (LBGs),
classified in different bins of SFR. To compare our models to observations, we first note that
all LBGs in the sample of \citet{Basu2012} fulfil the condition of high sSFR (see Sect.~\ref{res2}),
which ensures that their X-ray emission is dominated by HMXB populations. For these galaxies,
we need not model any further selection effects, because there is no threshold luminosity
(all the galaxies add their luminosities to the stack). Hence, the observational data are directly
comparable to the mean luminosity of galaxies of the intrinsic populations, classified 
in the same way as \citet{Basu2012} did. These authors report the galaxy luminosities in
the rest-frame $2-10\,{\rm keV}$ band. Hence, we transform our values of $L_{\rm X}$
to this luminosity range by assuming a photon index $\Gamma = 1.7$ for the X-ray spectra of individual binaries.

\subsection{Determination of the free parameters}
\label{free_par}

As stated in the Introduction, the goal of this work is to investigate the role played
by the chemical abundances  in the determination of
the number and  X-ray luminosity of HMXBs in star forming
galaxies. Our models explore the case
in which these properties are independent of the metallicity of the  HMXB progenitors (M0),
and two different kind of chemical dependences.
 A mild increase of the number of HMXBs
with decreasing metallicity by half an order of magnitude from $Z = 0.02$ to $Z = 0.0001$ 
as predicted by \citet{Belczynski2004a} is considered in M1. Models M2
and M3 assume a dependence of the rates and  a strong increase in the mean X-ray
luminosity of the stellar sources 
for very low metallicities,  $Z < 0.0003$.

We define three metallicity bins around the reference abundances of Belczynski et al.'s model: [0,0.0003], [0.0003,0.005]
and [0.005,0.04]. The number of sources are obtained by applying the scheme explained in Section 3.1 according to
the metallicity of the stellar populations. 
We do not include binary systems with $Z  > 0.04$ since the number and their effect in the total X-ray luminosity of the simulated galaxies
is negligible.
For M0 and M1, we adopt $L_{\rm max} = 10^{40} {\rm erg\, s}^{-1}$, regardless of the
metallicity. For M2 and M3, we increase arbitrarily 
$L_{\rm max}$ to  $10^{42} {\rm erg\, s}^{-1}$ only for sources in the low metallicity bin
(see Table~\ref{tab_parameters}).  The two order of magnitude
  higher $L_{\rm max}$ has been adopted to clearly enhance the
  possible effects
of a metallicity dependence.

 For each simulated galaxy at $z=0$, we generate its HMXB population according
to the metallicities and ages of its stellar populations, and consider detectability effects for nearby galaxies
as described in Sect.~\ref{res2}. Our models include two free parameters: a renormalization 
factor $\eta$, and the minimum X-ray luminosity of HMXBs $L_{\rm min}$, which will be chosen by confronting the
models with observational results of \citet{Mineo2012}. These observations provide  the number
of sources and total X-ray luminosity of the host galaxies as a
function of the SFR. 
Models M2 and M3 arise as  the result of preferentially matching either the
observed number or
the X-ray luminosity of the sources.

\begin{figure}
\centering
\includegraphics[width=0.4\textwidth]{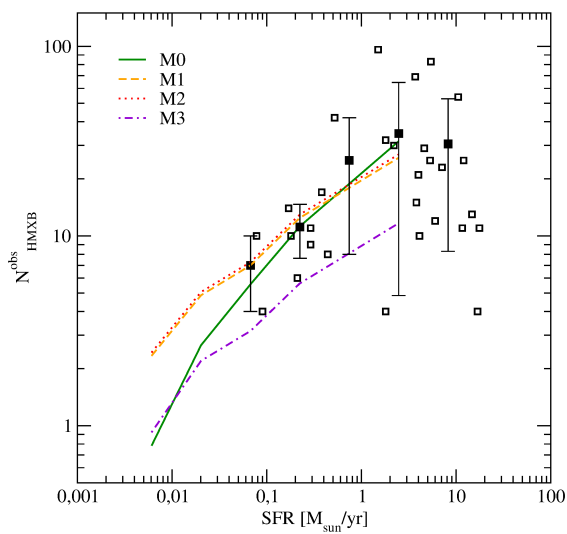}
\caption{Observed number of HMXBs in nearby, star-forming galaxies, as a function of their SFR.
Empty squares represent individual galaxies from the sample of \citet{Mineo2012};
filled ones are the mean values in each SFR bin. Model predictions for the mean value 
of $N_{\rm HMXB}^{\rm obs}$ are plotted  for M0 (green, solid line),
M1  (orange, dashed line),  M2 (red, dotted line) and  M3 (violet,
dash-dotted line). All models are required to  match the data in
order to determine the best-fitting  
$L_{\rm min}$ and $\eta$. M1 and M2 present the same relation. For the
sake of clarity, the relation of M2 has been displaced slightly.}
\label{near_N_SFR}
\end{figure}

\begin{figure}
  \centering
  \includegraphics[width=0.4\textwidth]{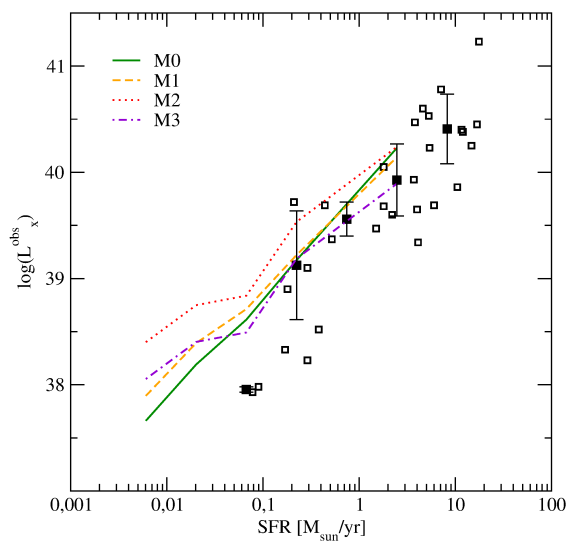}
\caption{Observed X-ray luminosity of nearby, star-forming galaxies, as a 
function of their SFR. Empty squares represent individual galaxies; filled
ones are the mean values in each SFR bin. Model predictions for the mean value 
of $L_{\rm X}^{\rm obs}$ are included for M0 (green, solid line), M1
(orange,dashed line),  
M2 (red, dotted line) and M3 (violet, dash-dotted line). All models
are required to 
match the data  in order to select the best-fitting $L_{\rm min}$ and
$\eta$ in each case. The dispersions of the model predictions 
are in the range 0.28--0.44~dex (except for M0, for which it is negligible), and have not been 
plotted for clarity. The slope of the $L_{\rm X}^{\rm obs}$--SFR relation decreases as the
chemical dependence of the model becomes stronger due to the correlation between the SFR 
and the metallicity of galaxies.}
\label{near_Lx_SFR}
\end{figure}

In Figs.~\ref{near_N_SFR}
and \ref{near_Lx_SFR}, we show $N^{\rm obs}_{\rm HMXB}$ and $L^{\rm obs}_{\rm X}$
as a function of SFR for our model predictions and the observations of nearby galaxies of \citet{Mineo2012}.
All the explored models are required to provide a good fit to the
data in order to determine the free parameters. The resulting values for each model
are listed in Table~\ref{tab_parameters}. The match of the models to
the observed data has been done
for the range ${\rm SFR} \sim 0.1 - 2\ M_\odot\ {\rm yr}^{-1}$, since there are no galaxies with higher
SFRs that fulfil the condition ${\rm sSFR} > 10^{-10}\ {\rm yr}^{-1}$
in our simulated catalogues.
The dispersions of the model predictions 
are in the range 0.28--0.44~dex (except for M0, for which it is negligible).  These dispersions are in excellent
agreement with those measured by \citet{Mineo2012} ($\sim 0.4$ dex), suggesting that the internal chemical inhomogeneity of
galaxies might be responsible for the observed dispersion in the  $L_{\rm X}$-SFR relation.

%
When  chemistry-dependent XLFs are also considered,  we find
that choosing the free parameters to match the observed mean number of sources (M2) overestimates the observed
mean luminosity of galaxies, while matching the latter (M3) slightly
underestimates the former. 
Unfortunately, the large error bars of the observed data prevent us to prefer one
over the other. Hence, we study  both of them 
to explore the consequences of a better fit to either observable. More
precise observations will certainly contribute to make a better selection of
the free parameters.

\section{HMXBs across cosmic time} \label{redshift_g}

\begin{figure*}
  \centering
  \includegraphics[width=0.35\textwidth]{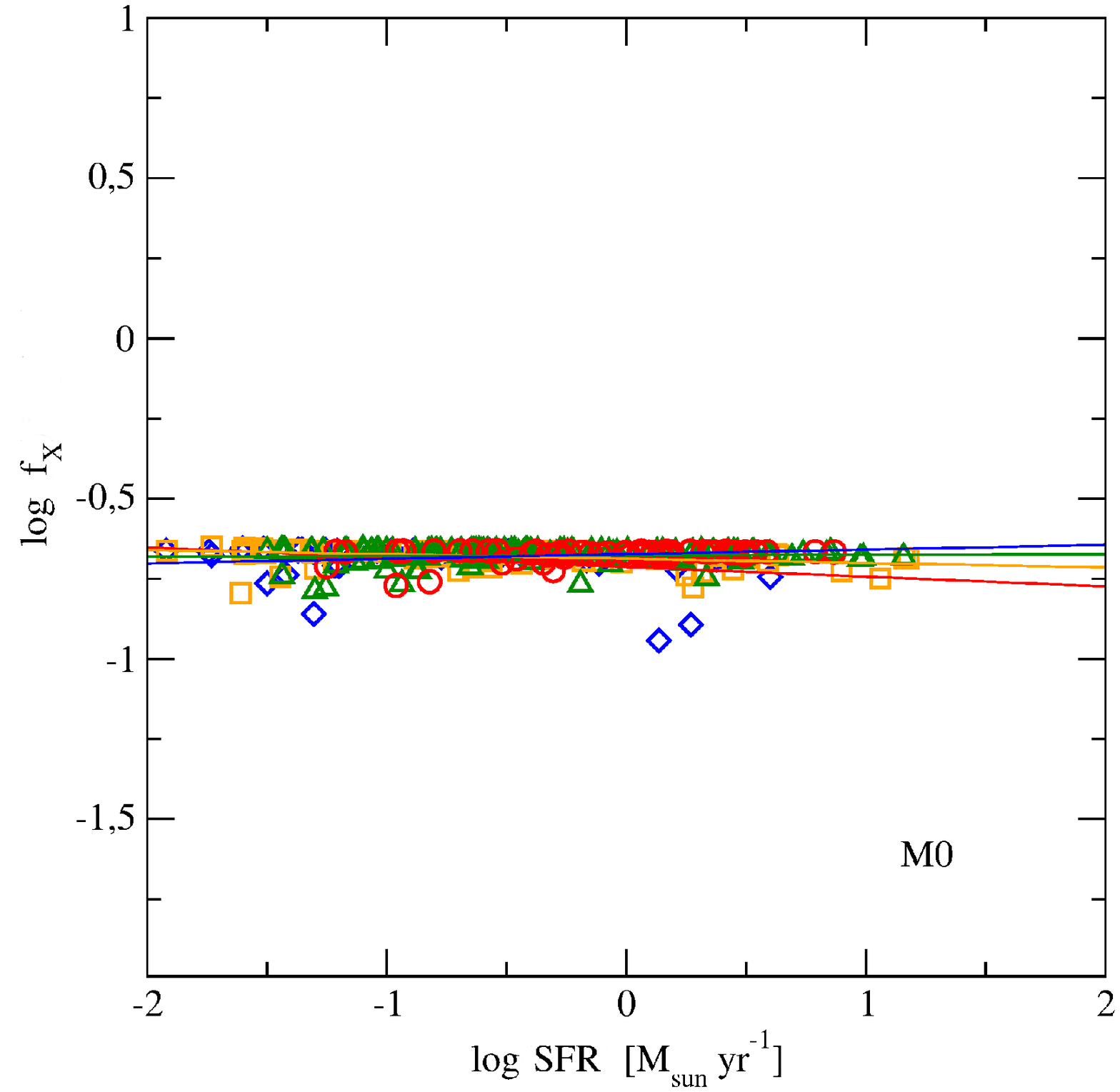}
   \includegraphics[width=0.35\textwidth]{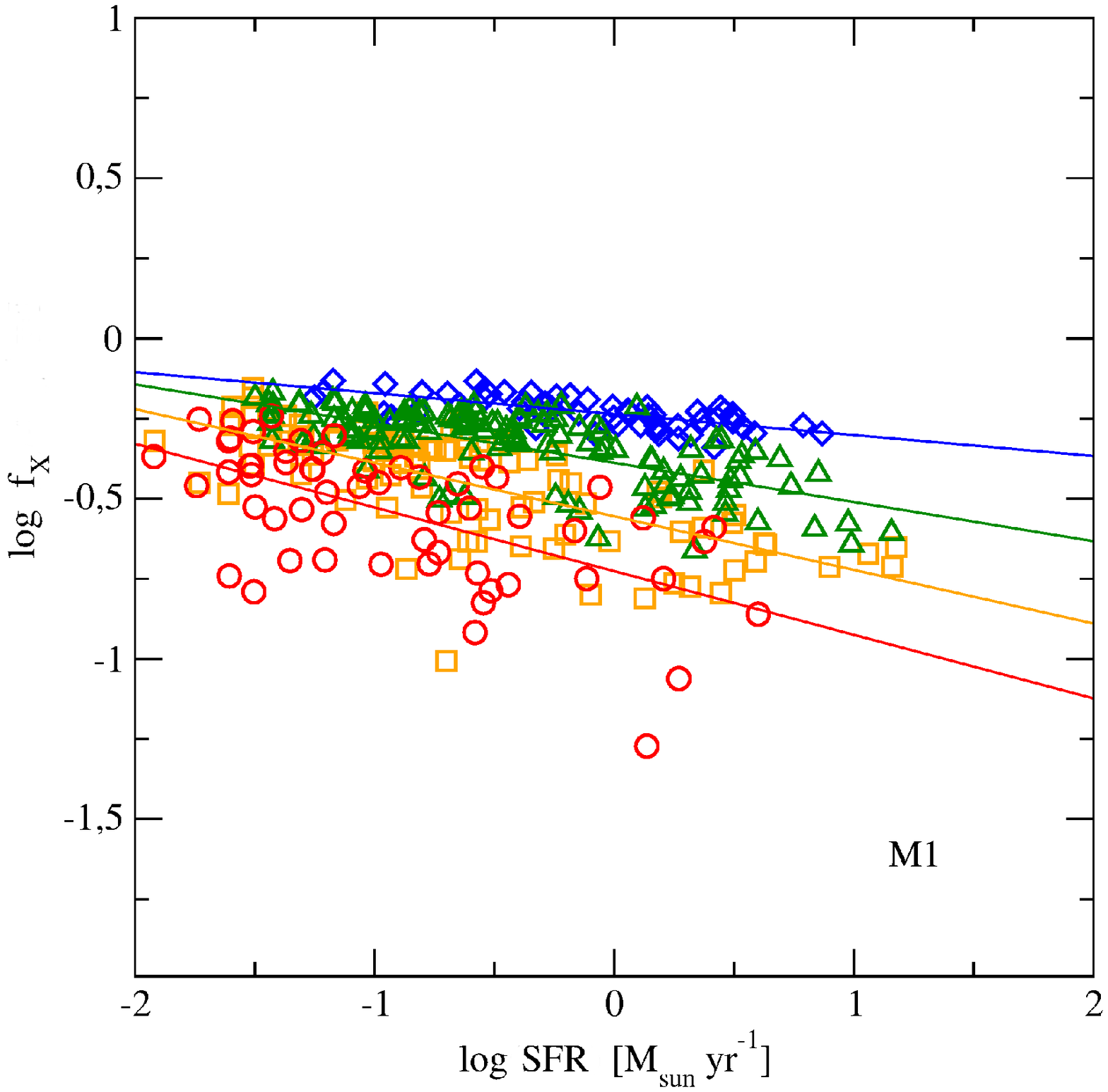}
  \includegraphics[width=0.35\textwidth]{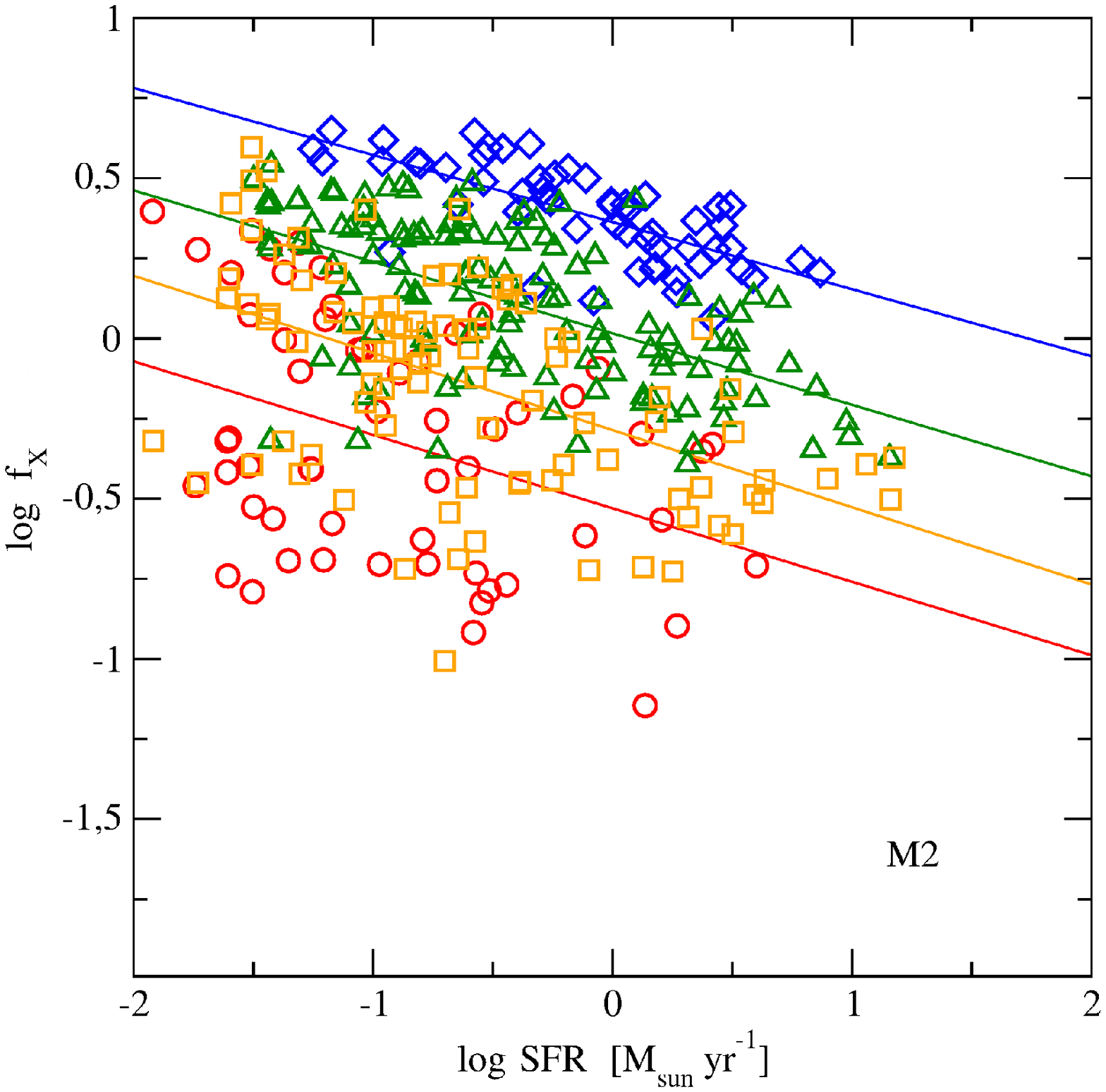}
    \includegraphics[width=0.35\textwidth]{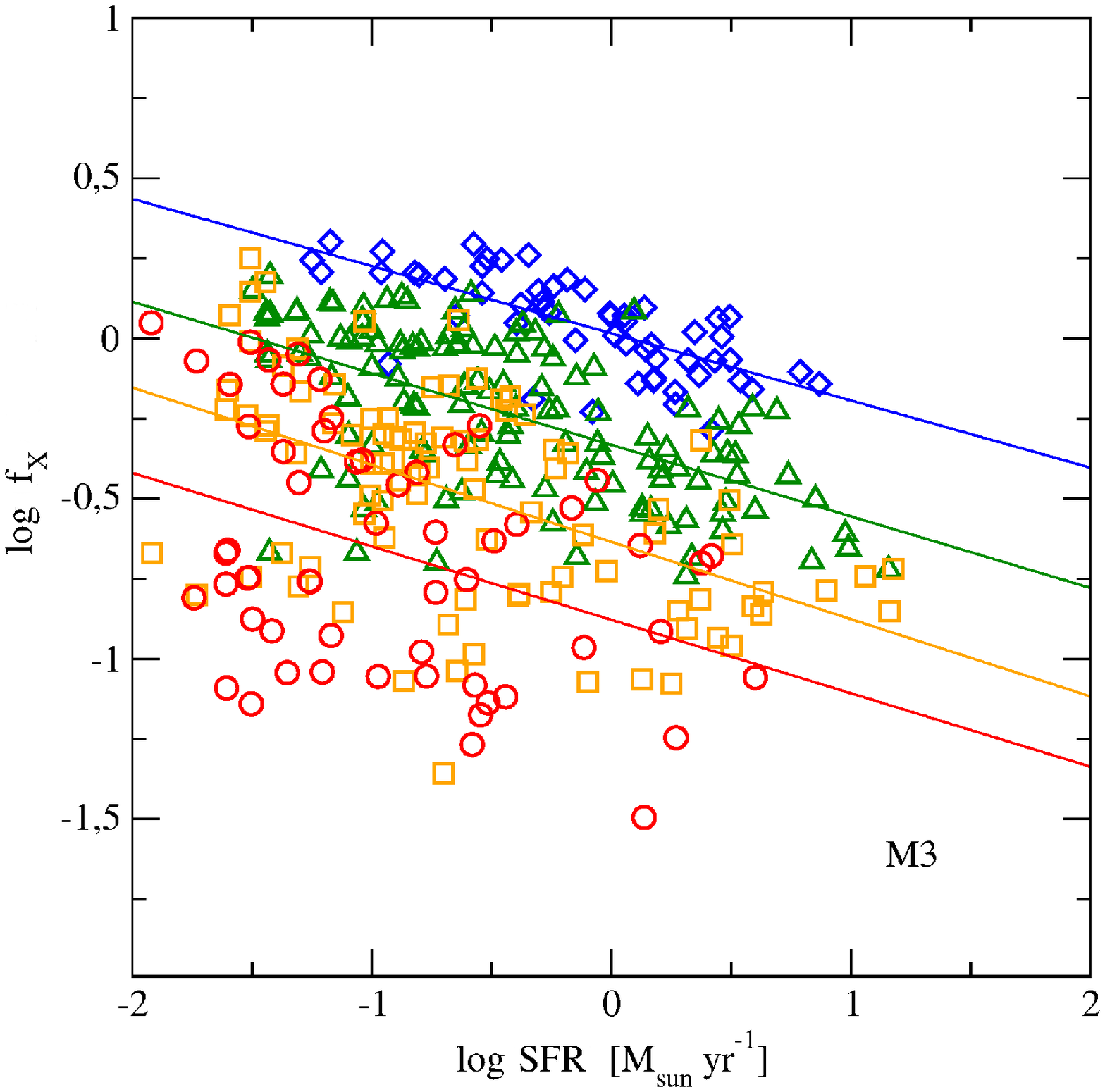}
\caption{The normalised ratio of the X-ray luminosity of galaxies to their SFR ($f_{\rm X}$) as a 
function of the latter, for four analysed models. The evolution of this relation is shown for  $z = 3.5$ (blue diamonds),
$z=2$ (green triangles), $z=1$ (orange squares) and $z= 0.3$ (red circles). The solid lines represent the best 
linear fits to the models at each redshift. At fixed redshift, chemistry-dependent models show a 
decrease of $f_{\rm X}$ with SFR, due to the correlation between the mean metallicity and the SFR of 
galaxies. The same models present an increase in $f_{\rm X}$ with redshift at fixed SFR, due to the
chemical evolution of the Universe. They also show a large dispersion in $f_{\rm X}$, due to the chemical 
inhomogeneities within each galaxy. All these effects are stronger in M2 and M3
(which include metallicity-dependent of both the  rates and the
luminosities of the HMXBs) than in M1 (which includes only metallicity-dependent HMXB rates),
and are absent in M0 (which includes no metallicity dependence).
However, if the data is not analysed as a function of redshift, these
parameters determine a relation consistent with a flat $f_{\rm X}$--SFR with a
large dispersion.}
\label{lx-sfr-z-1}
\end{figure*}

To investigate the evolution of the $L_{\rm X}$--SFR relation, we
analyse  $f_{\rm X}$ 
as function of redshift (see definition in Section~\ref{intro}).
 In Fig.~\ref{lx-sfr-z-1},
we plot  $f_{\rm X}$ predicted by our four models as a function of  SFR, for different redshifts in the range $z = 0.3 - 3.5$. 
As expected, for M0 the value of $f_{\rm X}$ is independent of
both SFR and redshift since by 
construction it takes a value of $\sim 0.2$, similar to that observed in the local Universe. 
This is due to the lack of any metallicity dependences so that  all galaxies produce
 X-ray luminosities proportional to their SFRs, with the same
 proportionality independently of redshift. The very small dispersion detected for  $f_{\rm X}$  
originates in the upper metallicity limit for binary systems to be able to produce 
HMXBs. As explained in the previous section, no  HMXB sources are produced for $Z > 0.04$. 
This introduces a small dispersion in the $f_{\rm X}$--SFR relation
 which gets higher as the metallicity of the interstellar medium from which stars form increases with decreasing redshift.

The effects of the chemical dependence included in M1 can be clearly
seen in Fig.~\ref{lx-sfr-z-1} which shows how $f_{\rm X} $ increases  with increasing
redshift, at a given SFR.
The correlation between the mean metallicity and SFR of galaxies, in the sense that 
high-SFR objects tend to be  more enriched than low-SFR ones, makes $f_{\rm X}$ to decrease with SFR 
at given redshift. 
These effects are consequence of the chemical evolution of the
Universe so  that  galaxies tend to be less enriched at high redshift, 
producing more HMXBs. 
The effects discussed above are mild for M1 
because they are only driven by the increase in the HMXB production
rate with decreasing metallicity.
Stronger trends are detected in models M2 and M3 
because of   the increase in both 
the HMXB production rate and their X-ray  luminosity at low
metallicities. 

\begin{figure}
\centering
\includegraphics[width=0.44\textwidth]{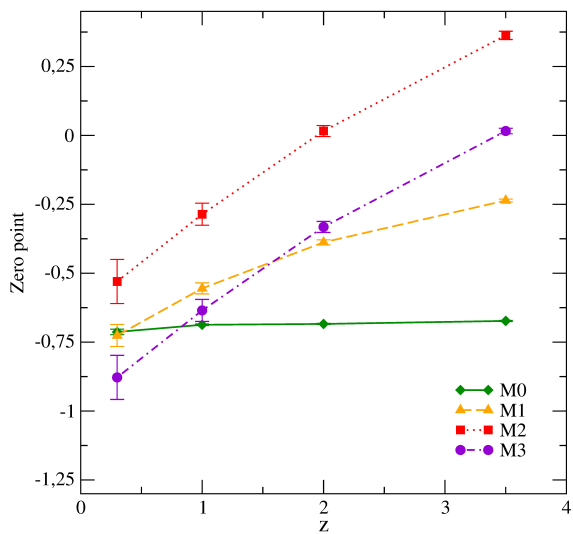}
\caption{Zero point of the best linear fit to the logarithmic $f_{\rm X}$--SFR
relation for each model (M0, green solid line and diamonds; M1, orange dashed line and 
triangles; M2, red dotted line and squares; M3, violet dot-dashed line and circles), as a
function of redshift. The zero point of the fit increases with redshift due to the 
chemical enrichment of the Universe. This increase is larger in models with a stronger chemical 
dependence (M2 and M3) and almost null in that with no chemical dependence (M0).}
\label{zeropoint}
\end{figure}

\begin{figure}
\centering
\includegraphics[width=0.44\textwidth]{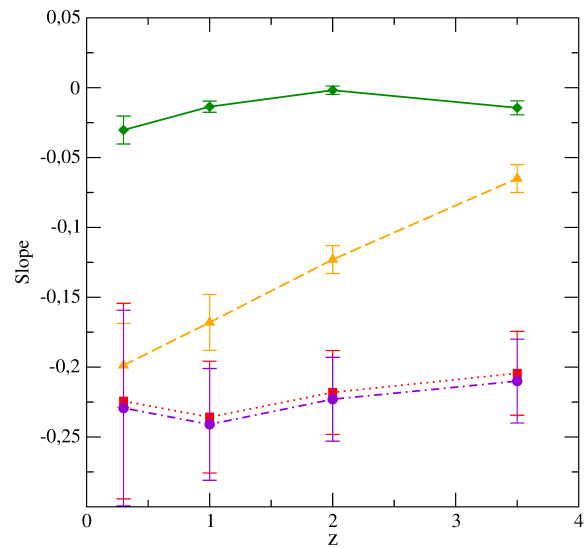}
\caption{Slope of the best linear fit to the logarithmic $f_{\rm X}$--SFR relation for
M0 (green diamonds and solid line),  M1 (orange triangles and dashed
line), M2 (red squares and
 dotted line) and  M3 (violet circles and  dot-dashed line), as a function of 
redshift. The absolute value of the slope of the fit is larger in models with a stronger chemical dependence (M2 and M3),
and almost null in that with no chemical dependence (M0), as expected
because of  the correlation between the SFR and
the mean metallicity of galaxies. M2 and M3 determine the same
relation; for the sake of clarity, we shifted slightly M2.}
\label{slope}
\end{figure}

\begin{figure}
\centering
\includegraphics[width=0.44\textwidth]{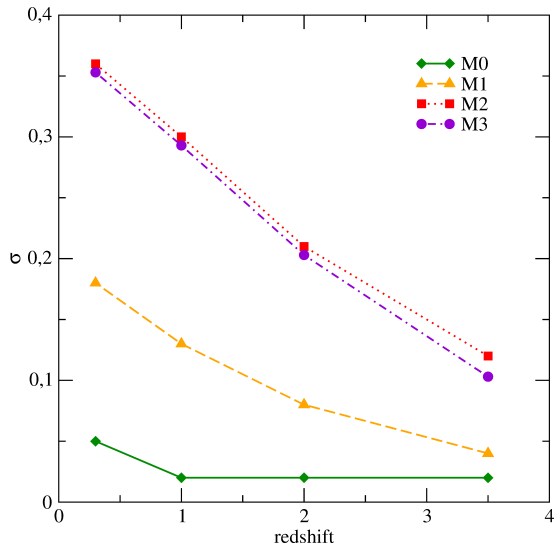}
\caption{Dispersion of the best linear fit to the logarithmic $f_{\rm X}$--SFR relation for 
each model (M0, green solid line and diamonds; M1, orange dashed line and triangles; M2,
red dotted line and squares; M3, violet dot-dashed line and circles), as a function of redshift.
The dispersion of the fit is larger in models with a stronger chemical dependence 
(M2 and M3) and almost null in that with no chemical dependence (M0), due to the chemical 
inhomogeneity of the star formation in galaxies. The decrease of the dispersion with 
redshift arises because at high redshift this inhomogeneity is smaller. M2 and M3 
determine the same relation; for sake of clarity, we shifted M2 slightly.}
\label{dispersion}
\end{figure}

It is interesting to note that trends shown  in Fig.~\ref{lx-sfr-z-1} are detected only when galaxies are
grouped  as a function of redshift. If no redshift information is
taken into account,  the mean trend is consistent with a  flat $f_{\rm  X}$--SFR relation,
with a large  dispersion as  the only observable effect of metallicity.
We quantify the evolution of the $f_{\rm X}$--SFR relation with cosmic time
by performing  linear regression fits to the simulated data at different redshifts. 
Table~\ref{ajuste_modelos} summarizes the best-fitting parameters, and
Fig.~\ref{zeropoint}--\ref{dispersion} show the evolution with redshift of the zero point,
slope, and dispersion of the fits, respectively.

 As can be seen from Fig.~\ref{zeropoint}, between $z \sim 0.3$ and $z \sim 3.5$, the
zero point  varies by $\sim 0.5\ {\rm dex}$ for M1 and by $\sim 1\ {\rm dex}$
for M2 and M3 (no evolution is seen for M0 as expected). As explained above, this variation is caused by 
the global chemical enrichment of the galaxies as they evolve. At low
redshift,  galaxies tend to be more chemically enriched, leading to lower $f_{\rm X}$ values. M2 and M3 show the same zero-point
trend with redshift, but the actual values differ by $\sim 0.4\ {\rm dex}$. This is consistent
with the construction process  as M2 was forced to fit the number of sources in nearby galaxies, 
hence overestimating the galaxy luminosities, and M3 was forced to 
fit the X-ray  luminosities themselves.

Fig.~\ref{slope} shows the evolution of the slope of  $f_{\rm X}$--SFR
relations with cosmic time.  For M0, the slope of the $f_{\rm X}$--SFR
relation is almost null at all redshifts, as expected. For M1, we
detect   a variation from $\sim -0.1$ to $\sim -0.2$ in the
analysed redshift range.
The negative value of the slope arises from
the correlation between the mean galaxy metallicity and SFR. Because of this correlation,
low-SFR galaxies tend to have preferentially low metallicities, hence chemistry-dependent
models produce higher $f_{\rm X}$ values for them. The variation of the slope with redshift
is due to the evolution of this correlation. The metallicity difference between low-SFR and 
high-SFR objects increases with decreasing redshift, as a consequence of their different 
chemical evolution. The absolute value of the (negative) slope of the $f_{\rm X}$--SFR 
relation follows this behaviour, which is clearly seen in Fig.~\ref{slope}. For M2 and M3,
the value of the slope at any redshift is lower than that of M1, due to the stronger dependence
of HMXB properties on metallicity in these models. 
The lack of 
evolution of the slope with redshift in M2 and M3 can be understood by comparison with M1,
recalling that M2 and M3 also include a strong increase in the luminosity of the most
metal-poor stellar populations. For low redshifts there are few such stellar populations, and
they are present mainly in low-SFR galaxies. Hence, M2 and M3 produce a small increase 
(with respect to M1) in the luminosities of these galaxies, making the slope of the $f_{\rm X}$--SFR
relation slightly more negative than in M1. For high redshifts, very low metallicity stellar
populations are ubiquitous. The fraction of the star formation that proceeds at these metallicities
increases as the SFR decreases, due to the SFR-metallicity correlation. The high X-ray luminosities
of these stellar populations amplify the effect of the variation of the number of HMXBs with metallicity.
Therefore, the slope of the $f_{\rm X}$--SFR relation is  much more negative
in M2 and M3 than in M1 for these redshifts. As a consequence, in M2 and M3 the slope 
attains similar values.
Intermediate values of $L_{\rm max}(Z_1)$ between the two tested maximum limits produce 
 relations in between the ones shown in   Fig.~\ref{slope}. Hence, if the slope of 
$f_{\rm X}$--SFR could be measured observationally as a function of redshift, it could provide information
not only on the existence of a metallicity-dependence of the X-ray luminosity of the   HMXBs but also on the maximum value 
it could attain.

The dispersion of the  $f_{\rm X}$--SFR relation for the simulated
HMXBs also change with
cosmic times. As it has been discussed above, the dispersion is
originated by the chemical inhomogeneity of stellar
populations in the simulated  galaxies, which is then reflected in the generated
HMXB sources in those chemistry-dependent models.
As a consequence,  
for M0 the dispersion is  negligible at any redshift (Fig.~\ref{dispersion}).
For M1, the dispersion decreases from $\sim 0.1$ at low redshifts to 
$\sim 0.04$ at $z \sim 3.5$, because for higher redshifts galaxies are
both less chemically enriched and
have lower metallicity dispersions. The same trend is seen in M2 and M3
showing an increase from $\sim 0.1$ at $z \sim 3.5$ to  $\sim 0.35$ at $z \sim 0.3$.
The higher dispersions  detected in M2 and M3 compared to those in M1
are due to the fact that  the formers also 
include the dependence on metallicity of the X-ray luminosities of
HMXBs as already mentioned.

The trend of $f_{\rm X}$
to increase with redshift has been reported to be marginal by \citet{Cowie2012}.
However, based on a larger set of observations, \citet{Basu2012} found a clear evolution of 
$f_{\rm X}$ with redshift, which can be parametrized as
$\log f_{\rm X} = (0.93 \pm 0.07) \log (1+z) - (0.35 \pm 0.03) \log {\rm SFR} - (0.74 \pm 0.03)$. 
We note that this relation includes not only the increase of $f_{\rm X}$ with redshift but also
its decrease with SFR. Taken as a function of SFR, the slope and zero point of this parametrization,
and the variation of the latter with redshift, are in reasonable agreement with the corresponding
predictions of our chemistry-dependent models M1 and M3.

\begin{figure*}
\centering
\includegraphics[width=0.85\textwidth]{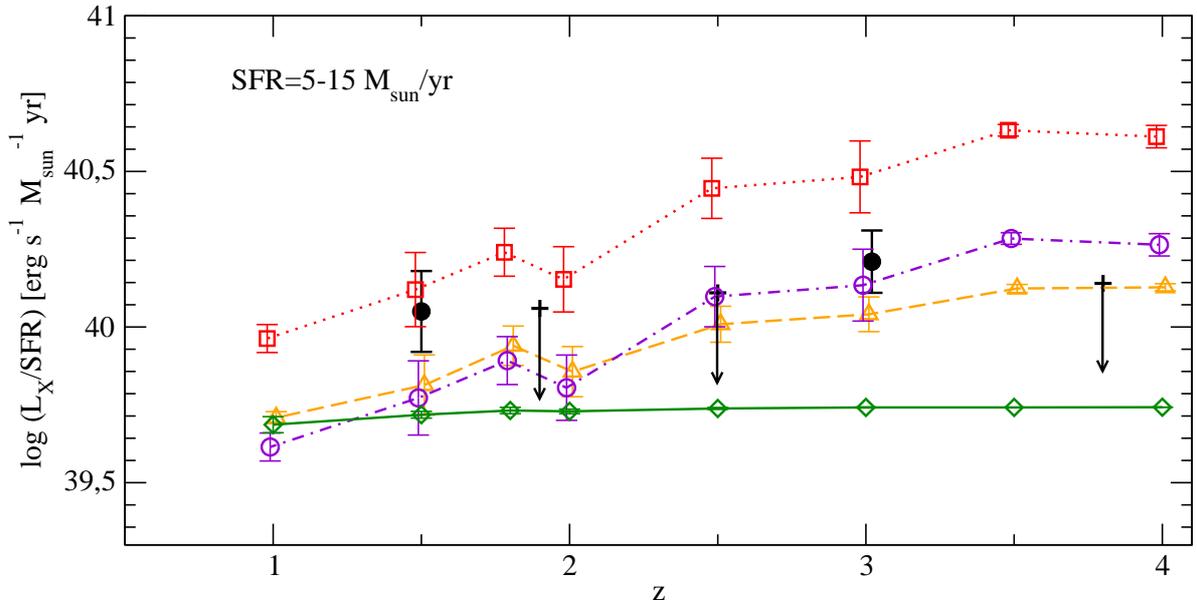}
\caption{Evolution of the ratio of the mean X-ray luminosity of 
galaxies to their SFR with redshift, for galaxies with ${\rm SFR} \in [5,15] M_{\odot}\ {\rm yr}^{-1}$. 
The prediction of our models (M0, green solid line and diamonds; M1,
orange dashed line and triangles; 
M2, red dotted line and squares; M3, violet dot-dashed line and circles) are shown against 
observed data (black circles and arrows) of \citet{Basu2012}. It is
clear from the plot,  and confirmed by Bayesian analysis, that only chemistry-dependent models reproduce the
trend seen in the observations.}
\label{BZ-comp}
\end{figure*}

To compare our models with observations of 
\citet{Basu2012} in more detail,  we use those galaxies with SFR in the range $5-15
M_{\odot}\ {\rm yr}^{-1}$. 
This range was chosen because both models and observations are well
represented within it. 
As can be seen  from  Fig.~\ref{BZ-comp},  only 
chemistry-dependent models can describe the trend seen in the
observations. Model M0
predicts $f_{\rm X}$ values far lower than those observed while M2 is at odds with the 
upper limits of \citet{Basu2012}. Only M1 and M3 seem to be consistent with the data.
This is confirmed by a Bayesian analysis; the posterior probabilities of the models 
given the data (including upper limits) are non-negligible only for
models M1 and M3.
 M2 shows an  excess of X-ray luminosity as expected because of the
higher number of sources predicted by this model (see Section 3).
These results suggest that the metallicity dependence is needed to explain the data, 
although   more observations and larger simulated volumes are clearly
required to improve this comparison.

\begin{table}
 \centering
 \caption{Parameters of the best linear regression
 $\log f_{\rm X} = A \log {\rm SFR}[M_\odot\ {\rm yr}^{-1}] + B$ for models M0--M3 at redshifts $z = 0.3-3.5$.}
\label{ajuste_modelos}
  \begin{tabular}{c c c c c}
  \hline \hline
Model & $z$ & $A$ & $B$ & $\sigma$\tablefootmark{*} \\ \hline
M0 & 0.3 & $ -0.03  \pm 0.01$  & $-0.71 \pm 0.01$  & 0.05 \\
M0 & 1.0 & $ -0.010 \pm 0.004$ & $-0.687 \pm 0.003$ & 0.02 \\
M0 & 2.0 & $ \phantom{-}0.002 \pm 0.003$ & $-0.677 \pm 0.003$ & 0.02 \\
M0 & 3.5 & $ \phantom{-}0.014 \pm 0.005$ & $-0.673 \pm 0.003$ & 0.02 \\ \hline
M1 & 0.3 & $-0.19 \pm 0.03$  & $-0.73 \pm 0.05$  & 0.18 \\
M1 & 1.0 & $-0.17 \pm 0.02$  & $-0.56 \pm 0.02$  & 0.13 \\
M1 & 2.0 & $-0.12 \pm 0.01$  & $-0.388 \pm 0.009$ & 0.08 \\
M1 & 3.5 & $-0.065 \pm 0.009$ & $-0.236 \pm 0.005$ & 0.04 \\ \hline
M2 & 0.3 & $-0.22 \pm 0.07$  & $-0.53 \pm 0.08$  & 0.35 \\
M2 & 1.0 & $-0.24 \pm 0.04$  & $-0.28 \pm 0.04$  & 0.29 \\
M2 & 2.0 & $-0.22 \pm 0.03$  & $-0.00 \pm 0.02$  & 0.20 \\
M2 & 3.5 & $-0.21 \pm 0.03$  & $ \phantom{-}0.40 \pm 0.01$  & 0.11 \\ \hline
M3 & 0.3 & $-0.23 \pm 0.07$  & $-0.88 \pm 0.08$  & 0.35 \\
M3 & 1.0 & $-0.24 \pm 0.04$  & $-0.63 \pm 0.04$  & 0.29 \\
M3 & 2.0 & $-0.22 \pm 0.03$  & $-0.33 \pm 0.02$  & 0.20 \\
M3 & 3.5 & $-0.21 \pm 0.03$  & $-0.02 \pm 0.01$  & 0.20 \\ \hline
\end{tabular}
\tablefoot{
 \tablefoottext{*}{Dispersion of the $f_{\rm X}$-SFR relation.}
}
\end{table}

\section{Discussion and Summary}
\label{con}

 Motivated by recent studies which suggest that both
the number of HMXBs and their X-ray luminosity might be higher in low-metallicity 
stellar systems \citep{Belczynski2004a,Dray2006,Linden2010,Mirabel2011a}, we explored the 
consequences of these hypotheses for the X-ray emission of star-forming galaxies,
and its cosmological evolution. For this purpose, we develope a  model 
to generate  HMXBs which is  applied to   galaxy catalogues constructed from 
hydrodynamical cosmological simulations. These simulations include a
self-consistent treatment of  the chemical 
evolution of baryons, and consequently, they provide the ages and
metallicities of the stellar populations as galaxies form and evolve.
As a function of time, each galaxy is described as mixture of stellar
populations with different metallicities and ages. By using our HMXBs
we can follow the formation of these events at different stages of
evolution of the Universe and confront them with observations to
constrain the free parameters of the models.   

We first confront our models with observations in nearby
galaxies \citep{Mineo2012} to estimate the free parameters. Then, we  
apply them to
investigate the variations of the HMXB properties across cosmic
time. We explore a non-metallicity dependent model and three other
ones with metallicity dependences in the production rate and X-ray luminosities.

We detect a significant dispersion in the $L_{\rm X}$--SFR relation
for our simulated local-Universe galaxies  in the range $\sim
  0.28-0.44$~dex, comparable to the $\sim 0.4$ dex reported by
  \citet{Mineo2012}. Hence, our results suggest that the internal
  metallicity dispersion of galaxies combined with the metallicity
  dependence of the X-ray sources
might provide  the physical origin for the observed dispersion in the $L_{\rm X}$--SFR relation.

We  explored the cosmological evolution of the ratio between the X-ray
luminosity of galaxies and their SFR, parametrized by the factor
$f_{\rm X}$. The confrontation of our models with the observations of 
\citet{Basu2012} favours models with metallicity
dependence, and rejects those in which  metallicity plays a negligible
role. However, the nature of this dependence cannot be 
determined  by using these observations. Both a dependence of the HMXB production rate on metallicity (M1), and this one 
plus a metallicity-dependent HMXB luminosity (M3) fit the data as well. 
More precise measurements of  $f_{\rm X}$--SFR relation as a function of redshift would help us to establish its  nature.

 In fact, three  main trends are detected in our chemistry-dependent models,
   which might  be used by observers to test the existence of a
   chemical dependence in the rate of production or in the luminosities of HMXBs:

\begin{itemize}

\item  At  given redshift, a decrease of $f_{\rm X}$ with the SFR of
  galaxies is detected  due to the correlation between the mean metallicity of galaxies
and their SFR. This correlation makes low-SFR galaxies have lower metallicity and hence, higher $f_{\rm X}$
that high-SFR ones. Our models predict a weak decrease if a
metallicity dependence  for the production rates is adopted  ($\sim 0.15$ dex) 
and a stronger one if this dependence is extended
to the X-ray luminosities ($\sim 0.25$ dex). 


\item For 
galaxies with similar SFRs, $f_{\rm X}$ should decrease with decreasing redshift because galaxies evolve to higher 
mean metallicities for decreasing redshift. Again, the level of
decrease is predicted to be directly related to the metallicity
dependence: the higher it is, the larger the change with redshift.
In our  models, $f_{\rm X}$ 
increases by  $\sim 0.5\ {\rm dex}$ between $z \sim 0$ and $z \sim 3.5$ for a metallicity-dependent
HMXB rate, and by  $\sim 1\ {\rm dex}$ in the same redshift range if metallicity affects 
both the rates and  the X-ray luminosities of HMXBs.

\item The $f_{\rm X}$--SFR relation shows a dispersion which
 reflects the combined effects of the chemical evolution of the
 stellar populations  and the metallicity dependences of the X-ray
 sources.
 This dispersion should
decrease with increasing redshift since high-redshift galaxies tend to have
stellar populations with 
more homogeneous metallicity distributions. Our findings suggest that
the variation of the dispersion with redshift store information on the
nature of the metallicity dependence.

\end{itemize}
None of these three effects is predicted in a scenario with a
  negligible metallicity dependence of the properties of HMXBs. Hence
  the observational measurement of any of them would make a strong case for this dependence.

Our  results suggest that the evolution of the $f_{\rm X}$--SFR relation
 should be observed up to high redshift and low SFRs  in order to assess
the chemical dependence of the properties of HMXB populations. 
The study of the properties of HMXB populations 
through cosmic times can unveil their potential contribution to energy feedback, which is 
expected to play a critical role in the thermal
and ionization history of the Universe \citep[e.g.][]{Mirabel2011a,Justham2012,Fragos2013a,Jeon2013}.
In a future work we explore the effects that energy feedback from
HMBXs might have on the regulation of the star formation in early
Universe.


\section*{Acknowledgements}
We would like to thank Laura Sales for useful comments. MCA acknowledges support from the European Commission's Framework Programme 7, 
through the Marie Curie International Research Staff Exchange Scheme
LACEGAL (PIRSES--GA--2010--269264). Simulations and tests were run in Fenix
Cluster (IAFE) and  Hal Cluster (Universidad Nacional de Córdoba).
Part of this work was developed within Cosmocomp ITN and LACEGAL IRSES Networks of the European Community. 
This work was partially supported by PICT 2006-0245 and PICT 2011-0959 from Argentine ANPCyT, and PIP 2009-0305 awarded by Argentine CONICET.

\bibliographystyle{aa}

\def\apj{ApJ}
\def\apjs{ApJS}
\def\apjl{ApJ}
\def\aj{AJ}
\def\mnras{MNRAS}
\def\aa{A\&A}
\def\nat{Nature}
\def\araa{ARA\&A}
\def\aap{A\&A}

\bibliography{ArtaleMC_final}

\begin{thebibliography}{43}
\expandafter\ifx\csname natexlab\endcsname\relax\def\natexlab#1{#1}\fi

\bibitem[{{Artale} {et~al.}(2011){Artale}, {Pellizza}, \&
  {Tissera}}]{Artale2011a}
{Artale}, M.~C., {Pellizza}, L.~J., \& {Tissera}, P.~B. 2011, \mnras, 415, 3417

\bibitem[{{Basu-Zych} {et~al.}(2013){Basu-Zych}, {Lehmer}, {Hornschemeier},
  {Bouwens}, {Fragos}, {Oesch}, {Belczynski}, {Brandt}, {Kalogera}, {Luo},
  {Miller}, {Mullaney}, {Tzanavaris}, {Xue}, \& {Zezas}}]{Basu2012}
{Basu-Zych}, A.~R., {Lehmer}, B.~D., {Hornschemeier}, A.~E., {et~al.} 2013,
  \apj, 762, 45

\bibitem[{{Belczynski} {et~al.}(2010{\natexlab{a}}){Belczynski}, {Bulik},
  {Fryer}, {Ruiter}, {Valsecchi}, {Vink}, \& {Hurley}}]{Belczynski2010b}
{Belczynski}, K., {Bulik}, T., {Fryer}, C.~L., {et~al.} 2010{\natexlab{a}},
  \apj, 714, 1217

\bibitem[{{Belczynski} {et~al.}(2010{\natexlab{b}}){Belczynski}, {Dominik},
  {Bulik}, {O'Shaughnessy}, {Fryer}, \& {Holz}}]{Belczynski2010a}
{Belczynski}, K., {Dominik}, M., {Bulik}, T., {et~al.} 2010{\natexlab{b}},
  \apjl, 715, L138

\bibitem[{{Belczynski} {et~al.}(2004{\natexlab{a}}){Belczynski}, {Kalogera},
  {Zezas}, \& {Fabbiano}}]{Belczynski2004b}
{Belczynski}, K., {Kalogera}, V., {Zezas}, A., \& {Fabbiano}, G.
  2004{\natexlab{a}}, \apjl, 601, L147

\bibitem[{{Belczynski} {et~al.}(2004{\natexlab{b}}){Belczynski}, {Sadowski}, \&
  {Rasio}}]{Belczynski2004a}
{Belczynski}, K., {Sadowski}, A., \& {Rasio}, F.~A. 2004{\natexlab{b}}, \apj,
  611, 1068

\bibitem[{{Belczynski} {et~al.}(2008){Belczynski}, {Taam}, {Rantsiou}, \& {van
  der Sluys}}]{Belczynski2008}
{Belczynski}, K., {Taam}, R.~E., {Rantsiou}, E., \& {van der Sluys}, M. 2008,
  \apj, 682, 474

\bibitem[{{Chisari} {et~al.}(2010){Chisari}, {Tissera}, \&
  {Pellizza}}]{Chisari2010}
{Chisari}, N.~E., {Tissera}, P.~B., \& {Pellizza}, L.~J. 2010, \mnras, 408, 647

\bibitem[{{Cowie} {et~al.}(2012){Cowie}, {Barger}, \& {Hasinger}}]{Cowie2012}
{Cowie}, L.~L., {Barger}, A.~J., \& {Hasinger}, G. 2012, \apj, 748, 50

\bibitem[{{De Rossi} {et~al.}(2013){De Rossi}, {Avila-Reese}, {Tissera},
  {Gonz{\'a}lez-Samaniego}, \& {Pedrosa}}]{deRossi2013}
{De Rossi}, M.~E., {Avila-Reese}, V., {Tissera}, P.~B.,
  {Gonz{\'a}lez-Samaniego}, A., \& {Pedrosa}, S.~E. 2013, \mnras, 435, 2736

\bibitem[{{de Rossi} {et~al.}(2010){de Rossi}, {Tissera}, \&
  {Pedrosa}}]{deRossi2010}
{de Rossi}, M.~E., {Tissera}, P.~B., \& {Pedrosa}, S.~E. 2010, \aap, 519, A89

\bibitem[{{De Rossi} {et~al.}(2012){De Rossi}, {Tissera}, \&
  {Pedrosa}}]{deRossi2012}
{De Rossi}, M.~E., {Tissera}, P.~B., \& {Pedrosa}, S.~E. 2012, \aap, 546, A52

\bibitem[{{Dijkstra} {et~al.}(2012){Dijkstra}, {Gilfanov}, {Loeb}, \&
  {Sunyaev}}]{Dijkstra2011}
{Dijkstra}, M., {Gilfanov}, M., {Loeb}, A., \& {Sunyaev}, R. 2012, \mnras, 421,
  213

\bibitem[{{Dray}(2006)}]{Dray2006}
{Dray}, L.~M. 2006, \mnras, 370, 2079

\bibitem[{{Fabbiano}(2006)}]{Fabbiano2006}
{Fabbiano}, G. 2006, \araa, 44, 323

\bibitem[{{Feng} \& {Soria}(2011)}]{Feng2011}
{Feng}, H. \& {Soria}, R. 2011, \nar, 55, 166

\bibitem[{{Fragos} {et~al.}(2013{\natexlab{a}}){Fragos}, {Lehmer}, {Tremmel},
  {Tzanavaris}, {Basu-Zych}, {Belczynski}, {Hornschemeier}, {Jenkins},
  {Kalogera}, {Ptak}, \& {Zezas}}]{Fragos2012}
{Fragos}, T., {Lehmer}, B., {Tremmel}, M., {et~al.} 2013{\natexlab{a}}, \apj,
  764, 41

\bibitem[{{Fragos} {et~al.}(2013{\natexlab{b}}){Fragos}, {Lehmer}, {Naoz},
  {Zezas}, \& {Basu-Zych}}]{Fragos2013a}
{Fragos}, T., {Lehmer}, B.~D., {Naoz}, S., {Zezas}, A., \& {Basu-Zych}, A.
  2013{\natexlab{b}}, \apjl, 776, L31

\bibitem[{{Georgy} {et~al.}(2009){Georgy}, {Meynet}, {Walder}, {Folini}, \&
  {Maeder}}]{Georgy2009}
{Georgy}, C., {Meynet}, G., {Walder}, R., {Folini}, D., \& {Maeder}, A. 2009,
  \aap, 502, 611

\bibitem[{{Grimm} {et~al.}(2003){Grimm}, {Gilfanov}, \& {Sunyaev}}]{Grimm2003}
{Grimm}, H.-J., {Gilfanov}, M., \& {Sunyaev}, R. 2003, \mnras, 339, 793

\bibitem[{{Jeon} {et~al.}(2013){Jeon}, {Pawlik}, {Bromm}, \&
  {Milosavljevic}}]{Jeon2013}
{Jeon}, M., {Pawlik}, A.~H., {Bromm}, V., \& {Milosavljevic}, M. 2013, ArXiv
  e-prints

\bibitem[{{Justham} \& {Schawinski}(2012)}]{Justham2012}
{Justham}, S. \& {Schawinski}, K. 2012, \mnras, 423, 1641

\bibitem[{{Kaaret} {et~al.}(2011){Kaaret}, {Schmitt}, \& {Gorski}}]{Kaaret2011}
{Kaaret}, P., {Schmitt}, J., \& {Gorski}, M. 2011, \apj, 741, 10

\bibitem[{{Linden} {et~al.}(2010){Linden}, {Kalogera}, {Sepinsky}, {Prestwich},
  {Zezas}, \& {Gallagher}}]{Linden2010}
{Linden}, T., {Kalogera}, V., {Sepinsky}, J.~F., {et~al.} 2010, \apj, 725, 1984

\bibitem[{{Maiolino} {et~al.}(2008){Maiolino}, {Nagao}, {Grazian}, {Cocchia},
  {Marconi}, {Mannucci}, {Cimatti}, {Pipino}, {Ballero}, {Calura}, {Chiappini},
  {Fontana}, {Granato}, {Matteucci}, {Pastorini}, {Pentericci}, {Risaliti},
  {Salvati}, \& {Silva}}]{Maiolino2008}
{Maiolino}, R., {Nagao}, T., {Grazian}, A., {et~al.} 2008, \aap, 488, 463

\bibitem[{{Mineo} {et~al.}(2012){Mineo}, {Gilfanov}, \& {Sunyaev}}]{Mineo2012}
{Mineo}, S., {Gilfanov}, M., \& {Sunyaev}, R. 2012, \mnras, 419, 2095

\bibitem[{{Mirabel} {et~al.}(2011){Mirabel}, {Dijkstra}, {Laurent}, {Loeb}, \&
  {Pritchard}}]{Mirabel2011a}
{Mirabel}, I.~F., {Dijkstra}, M., {Laurent}, P., {Loeb}, A., \& {Pritchard},
  J.~R. 2011, \aap, 528, A149

\bibitem[{{Mosconi} {et~al.}(2001){Mosconi}, {Tissera}, {Lambas}, \&
  {Cora}}]{Mosconi2001}
{Mosconi}, M.~B., {Tissera}, P.~B., {Lambas}, D.~G., \& {Cora}, S.~A. 2001,
  \mnras, 325, 34

\bibitem[{{Nuza} {et~al.}(2007){Nuza}, {Tissera}, {Pellizza}, {Lambas},
  {Scannapieco}, \& {de Rossi}}]{Nuza2007}
{Nuza}, S.~E., {Tissera}, P.~B., {Pellizza}, L.~J., {et~al.} 2007, \mnras, 375,
  665

\bibitem[{{Pedrosa} {et~al.}(2014){Pedrosa}, {Tissera}, \& {De
  Rossi}}]{pedrosa2014}
{Pedrosa}, S.~E., {Tissera}, P.~B., \& {De Rossi}, M.~E. 2014, ArXiv e-prints

\bibitem[{{Pellizza} {et~al.}(2012){Pellizza}, {Artale}, \&
  {Tissera}}]{Pellizza2012}
{Pellizza}, L.~J., {Artale}, M.~C., \& {Tissera}, P.~B. 2012, in Gamma-Ray
  Bursts 2012 Conference (GRB 2012)

\bibitem[{{Power} {et~al.}(2013){Power}, {James}, {Combet}, \&
  {Wynn}}]{Power2013}
{Power}, C., {James}, G., {Combet}, C., \& {Wynn}, G. 2013, \apj, 764, 76

\bibitem[{{Power} {et~al.}(2009){Power}, {Wynn}, {Combet}, \&
  {Wilkinson}}]{Power2009}
{Power}, C., {Wynn}, G.~A., {Combet}, C., \& {Wilkinson}, M.~I. 2009, \mnras,
  395, 1146

\bibitem[{{Salpeter}(1955)}]{Salpeter1955}
{Salpeter}, E.~E. 1955, \apj, 121, 161

\bibitem[{{Scannapieco} {et~al.}(2005){Scannapieco}, {Tissera}, {White}, \&
  {Springel}}]{Scannapieco2005}
{Scannapieco}, C., {Tissera}, P.~B., {White}, S.~D.~M., \& {Springel}, V. 2005,
  \mnras, 364, 552

\bibitem[{{Scannapieco} {et~al.}(2006){Scannapieco}, {Tissera}, {White}, \&
  {Springel}}]{Scannapieco2006}
{Scannapieco}, C., {Tissera}, P.~B., {White}, S.~D.~M., \& {Springel}, V. 2006,
  \mnras, 371, 1125

\bibitem[{{Shtykovskiy} \& {Gilfanov}(2007)}]{Shtykovskiy2007}
{Shtykovskiy}, P.~E. \& {Gilfanov}, M.~R. 2007, Astronomy Letters, 33, 437

\bibitem[{{Springel}(2005)}]{Springel2005}
{Springel}, V. 2005, \mnras, 364, 1105

\bibitem[{{Springel} {et~al.}(2001){Springel}, {White}, {Tormen}, \&
  {Kauffmann}}]{Springel2001}
{Springel}, V., {White}, S.~D.~M., {Tormen}, G., \& {Kauffmann}, G. 2001,
  \mnras, 328, 726

\bibitem[{{Thielemann} {et~al.}(1993){Thielemann}, {Nomoto}, \&
  {Hashimoto}}]{Thielemann1993}
{Thielemann}, F.~K., {Nomoto}, K., \& {Hashimoto}, M. 1993, {Origin and
  Evolution of Elements}, 1st edn., ed. N.~{Prantzos}, E.~{Vangoni-Flam}, \&
  N.~{Cass} (Cambridge: Cambridge Univ. Press), p. 299

\bibitem[{{Tremonti} {et~al.}(2004){Tremonti}, {Heckman}, {Kauffmann},
  {Brinchmann}, {Charlot}, {White}, {Seibert}, {Peng}, {Schlegel}, {Uomoto},
  {Fukugita}, \& {Brinkmann}}]{Tremonti2004}
{Tremonti}, C.~A., {Heckman}, T.~M., {Kauffmann}, G., {et~al.} 2004, \apj, 613,
  898

\bibitem[{{Woosley} \& {Weaver}(1995)}]{Woosley1995}
{Woosley}, S.~E. \& {Weaver}, T.~A. 1995, \apjs, 101, 181

\bibitem[{{Zuo} \& {Li}(2011)}]{Zuo2011}
{Zuo}, Z.-Y. \& {Li}, X.-D. 2011, \apj, 733, 5

\end{thebibliography}

\IfFileExists{\jobname.bbl}{}
{\typeout{}
\typeout{****************************************************}
\typeout{****************************************************}
\typeout{** Please run "bibtex \jobname" to optain}
\typeout{** the bibliography and then re-run LaTeX}
\typeout{** twice to fix the references!}
\typeout{****************************************************}
\typeout{****************************************************}
\typeout{}
}


\end{document}